\documentstyle [11pt,epsf]{article}

\begin{document}
%
%
\newcommand{\mod}[0]{\mbox{ mod }}
\newcommand{\Abs}[1]{|#1|}
\newcommand{\Tr}[0]{\mbox{Tr}}
\newcommand{\EqRef}[1]{(\ref{eqn:#1})}
\newcommand{\FigRef}[1]{fig.~\ref{fig:#1}}
\newcommand{\Abstract}[1]{\small
   \begin{quote}
      \noindent
      {\bf Abstract - }{#1}
   \end{quote}
    }

\title{From chaotic to disordered systems -\\
a periodic orbit approach}

\author{Per Dahlqvist\\ 
Royal Institute of Technology\\
S-100 44 Stockholm, Sweden\\
}
\date{}

\maketitle

\begin{abstract}
We apply periodic orbit theory to a quantum billiard on a torus with a variable 
number $N$ of small circular scatterers distributed randomly. 
Provided these 
scatterers are much smaller than the wave length they
may be regarded as sources of 
diffraction.
The relevant part of the spectral determinant
is due to diffractive periodic orbits only.
We formulate this diffractive zeta function 
in terms of a $N \times N$ transfer matrix,
which is transformed to real form.
The zeros of this determinant can readily be computed. 
The determinant is shown to reproduce the full density of states
for generic configurations if $N\geq 4$.
We study the statistics exhibited by these spectra.
The numerical results suggest that the spectra
tend to GOE statistics as the number of scatterers increases 
for typical members
of the ensemble.
A peculiar situation arises for configurations with four scatterers and
$kR$ tuned to $kR=y_{0,1}\approx 0.899$, 
where the statistics appears to be perfectly Poissonian.

\end{abstract}

\section{Introduction}

Universal level statistics 
of classically chaotic systems
is an asymptotic (i.e. semiclassical) property of a spectrum.
For homogeneous systems,
such as billiards, it appears in the high energy limit of the spectrum.
Usually there is a preasymptotic regime where the statistics reflects
the characteristics of classical or  quantum origin 
specific to the system.

Many authors consider
classical diffusion and inhibition of quantum diffusion due to localization
as the key to understand nonuniversal features of spectra\cite{Casati}. 
In this paper we are
going to study the problem from a periodic orbit point of view.
Admittedly, periodic orbit theories based on
the semiclassical trace formula\cite{GutBook}
has not been very successful for studies of spectral statistics.
The exception is Berry result on the small $\tau$ limit of the form factor
\cite{BerryRig}.
This is derived  under the so-called {\em diagonal approximation},
an approximation of somewhat obscure validity\cite{PDdual}.

To simplify the
discussion of periodic orbit theories below we will
confine ourselves to dispersive
billiards consisting of one or several disks, all with  radius $R$, 
inside a rectangle 
or a torus.
We thus limit the number of relevant length scales
to essentially one.
One obvious condition for the trace formula to be applicable
is $kR\gg 1$, where $k=\sqrt{2mE}/\hbar$ is the wave number. 
Non universal effects are 
expected to be pronounced for intermediate values of $kR$,
and are thus
to be sought in the region where 
the validity of the trace formula is dubious.

Diffraction 
plays presumably a much more important role in
bound system than in open and the 
Gutzwiller formula has to be amended by sums over
diffractive periodic orbits\cite{vattay1,vattay2}.
However, it is not obvious how much the inclusion of diffractive orbits
will improve the situation.
The 
{\em geometric theory of
diffraction} is very successful for open systems, where it provides
small corrections, 
but its applicability
is questionable in the {\em penumbra} of disk scattering in the very forward
direction
\cite{Prim95,PDNaka}.
The basic problem is the interference
between neutral  and unstable orbits accumulating towards them. 
Their respective saddle points are not enough
separated and individual eigenstates can hardly be resolved by Berry-Keating
technique. 
This should hardly come as a surprise.
The failure of predicting individual eigenvalues in the semiclassical limit
has been expected from the very genesis of the trace formula.
The pessimism has partly fallen into oblivion since the success of the Berry-Keating
scheme \cite{SS91,BK90}.

The harsh moral of this discussion is that
it seems not an easy task to pursue periodic theories 
as a mean to study how universality
may emerge as $kR \rightarrow \infty$.

The situation turns out to be very different if the opposite limit
($kR\ll 1$) is considered.
The disks can now be considered as sources of s-wave diffraction.
The unstable orbits are replaced by
purely diffractive orbits.

In ref. \cite{DahlVattay} we studied the small $kR$ 
limit for the one disk case.
The limit is not only much easier to deal with than the opposite, 
but it is also much richer in behavior.
In particular the
$A_1$ subspace 
exhibit a wide range of level statistics in the diffractive region
which, due to symmetry effects, extend up to $kR \approx 4$.
For ($kR\approx 2.40$ and $kR\rightarrow 0$)
the statistics is Poissonian
and for ($kR\approx 0.899$) it is very close to GOE.
It approaches GOE properly first in the limit $kR\rightarrow \infty$.
One of the questions we are going to address is whether a GOE statistics can
be achieved by keeping $kR$ small, but  by increasing the number of scatterers,
and  distribute them  randomly over a torus. We will thus enter
the realm of disordered systems.

In Random Matrix Theories one studies ensemble averages of abstract models of disordered
systems whereas in Quantum Chaos one 
usually studies self-averages of chaotic systems.
There is a need to study how the predictions of 
Random Matrix Theories is realized in more concrete models
of disordered systems.
We have chosen study spectral statistics
for {\em individual} members of our disordered ensembles,
for various values of the parameters ($N$ and $kR$), although
nothing prevented us in principle from studying ensemble averages.
Any study of non-universality and approach to universality for a single
chaotic systems (utilizing self-averages of the spectrum) will
suffer from finiteness of the sample.
Some type of non-universal effects, like fine wiggles on the formfactor
proposed by \cite{Altshuler,Agam} may not relevant when
applied to a single systems\cite{Prange}, but this is not the kind of effect
we will be looking at.

The basic motivation in this paper is conceptual rather than physical, although
our studies have some bearing on disordered solids
and impurity scattering.
The system considered in this paper have obvious 
similarities with antidot lattices\cite{antidot}.
However, presentday antidot lattices 
are modeled by rather smooth potentials
and they
lie in the intermediate region
$kR \sim 2\pi $ (with $R$ suitably defined)
where periodic orbit theories are,
least to say, cumbersome.

The outline is as follows.
The eigenvalues are recognized as the zeros of the {\em spectral determinant},
which we will derive within the geometric theory of diffraction.
We will thus focus on the diffractive
determinant (or zeta function),
associated with periodic orbits
with at least one scattering on a diffractive object.
This is formulated in section \ref{s:def}.
This object has poles on the real axis so it has to be resummed
to tame the divergence caused by these poles. 
This is done in section \ref{s:resum}.
In section \ref{s:num} we discuss some numerical issues.
In section \ref{s:mean} we derive the man level density of 
zeros of the diffractive determinant.
In section \ref{s:stat} we compute spectra numerically
and study their statistics.

\section{The diffractive determinant}

\subsection{Derivation of the diffractive determinant}
\label{s:def}

In the {\em geometric theory of diffraction}\cite{Keller1,vattay1,vattay2}
the spectral determinant is split up into a product
\begin{equation}
\Delta(E)=\Delta_0(E)\cdot\Delta_G(E)\cdot\Delta_D(E) \ \ ,
\end{equation}
where $\Delta_0(E)$ corresponds 
to the mean level density, the geometric part 
$\Delta_G(E)$ is the Gutzwiller-Voros zeta function,
possibly amended with the neutral orbits.
We will be solely interested in the diffractive determinant $\Delta_D(E)$.
It has been derived for an non-diffractive system supplied with
$N$ small disks in ref \cite{DahlVattay}.

We thus assume the presence of
$N$ small diffractive objects 
located at ${\bf r}_k$, where $1\leq k \leq N$ whose diffraction constants
$d_k(E)$ 
do not depend on the scattering angle.
We introduce symbolic dynamics by enumerating the disk from $1$ to
$N$. The alphabet is now $\{k;1\leq k\leq N\}$.
The set $\Omega_D$ is defined as the set of
all primitive periodic sequences of symbols taken from this alphabet.
Note that the transition $\ldots k_i k_{i+1} \ldots$ does not correspond to
{\em one} trajectory from ${\bf r}_{k_i}$ to ${\bf r}_{k_{i+1}}$, 
as is usual in symbolic dynamics
but {\em all} trajectories. To clarify this
distinction we prefer to talk about {\em periodic symbol sequences} rather
than
{\em periodic orbits}.

The diffractive determinant (or zeta function) is now given by
\begin{equation}
\Delta_D(E)=\prod_{p \in \Omega_D} (1-t_p) \ \ .
\end{equation}
The weight $t_p$ is given by
\begin{equation}
 t_p= \prod_{i=1}^{n_p} d_{k_i} 
G_G({\bf r}_{k_{i-1}},{\bf r}_{k_{i}},E) \ \ ,  \label{eqn:tpdef}
\end{equation}
where $p=\overline{k_1 k_2 \ldots k_{n_p}}$.
and $k_0=k_n$.
$G_G({\bf r},{\bf r}',E)$ is the non diffractive
(i.e. the Green function for the original system, before the diffractive objects
have been inserted).
It can, in the semiclassical limit,
 be decomposed into a sum over all paths $(j)$ from
${\bf r}$ to ${\bf r}'$
\begin{equation}
G_G({\bf r},{\bf r}',E)=\sum_{j:q \mapsto q'} 
G_0^{(j)}({\bf r},{\bf r}',E) \ \ .  \label{eqn:pathsum}
\end{equation}
where $G_0$ is given by the usual van Vleck-Gutzwiller expressions \cite{GutBook}.


We now turn to our particular system;
a rectangle with sides $a$ and $b$ supplied with periodic boundary conditions,
with $N$ circular disks at positions ${\bf r}_k$ where $1\leq k \leq n$.
All disks have radius $R$ and their diffractive constants are given by
\begin{equation}
d(E)= -4i\frac{J_0(kR)}{H_0^{(1)}(kR)}  \  \ .  \label{eqn:d}
\end{equation}
which applies in the limit $kR\rightarrow 0$.
We will use as $G_0$ the free flight Greens function given by
\begin{equation}
G_0({\bf r},{\bf r}',E)
\equiv G_0({\bf r}'-{\bf r},E)
=-\frac{i}{4}H_0^{(1)}(k|{\bf r}-{\bf r}'|) \ \ ,
\end{equation}
instead of the of usual van Vleck-Gutzwiller, which is obtained by taking
the Debye approximation of the Hankel functions $H_0^{(1)}(z)$.
The geometric Green function now reads
\begin{equation}
G_G({\bf r},E)=\sum_{{\bf \rho}=(ma,nb)} G_0({\bf r}+\rho,E)  \  \ . 
\end{equation}
where the sum runs over all integer $m$ and $n$.

Due to a singularity of the Hankel function, this expression diverges if
${\bf r}_{12}\rightarrow 0$. We define a regularized geometric Green
$\tilde{G}_G({\bf r},E)$ function by subtracting this singularity.
The {\em diagonal} Green function from a disk to itself is now
\begin{eqnarray}
\tilde{G}_G({\bf r}=0,E)=-\frac{i}{4} 
\sum_{{\bf \rho}\neq{\bf 0}} H_0^{(1)}(k\rho) \label{eqn:Gdiag} \ \ ,
\end{eqnarray}
and the {\em off diagonal}
\begin{equation}
\tilde{G}_G({\bf r}\neq 0,E)=-\frac{i}{4} 
\sum_{{\bf \rho}} H_0^{(1)}(k|\rho+{\bf r}|)  \label{eqn:Goff} \ \ .
\end{equation}

Due to the multiplicative expression for the weights $t_p$ \EqRef{tpdef}
the diffractive determinant, or zeta function,
can be computed from the transfer matrix\cite{AAC,DasBuch}
\begin{equation}
T_{ij}=d(E) \cdot \tilde{G}_G({\bf r}_j-{\bf r}_i,E) 
\end{equation}
via
\begin{equation}
\Delta_D(E)=\mbox{det}(1-{\bf T})  \label{eqn:DeltaD} \ \ .
\end{equation}

\subsection{Making the sums converge and the determinant real}
\label{s:resum}

The sums \EqRef{Goff} and
\EqRef{Gdiag} diverge for real E and we will resort to the
Ewald summation technique as developed in ref.\ \cite{BerrySin}
in order to control the singularities.
This procedure transforms the
{\em diagonal} Green function to
\begin{equation}
\tilde{G}_G(0,E)= \frac{1}{ab} \sum_{{\bf g}=2\pi(m/a,n/a)} 
\frac{\exp (Q[ 1-{\bf g}^2/(2E)])}{2E-{\bf g}^2}-
\frac{1}{4\pi}Ei(Q)+\frac{i}{4}
\end{equation}
\[
-\frac{1}{4\pi}\sum_{{\bf \rho}\neq{\bf 0}} I(\frac{k}{2}|\rho|)
\equiv G^{(r)} (0,E)+\frac{i}{4}  \ \ ,
\]
where $I(x)$ is defined by the integral
\begin{equation}
I(x)=\int_{\log (x/Q)}^{\infty}\exp(-2x \sinh \xi ) d\xi 
\ \ .  \label{eqn:Ixdef}
\end{equation}
These expressions are identical to those in \cite{BerrySin}, we
just keep  $a$ and $b$ as free parameters.

The {\em off diagonal} Green function  \EqRef{Goff}
is after resummation
\begin{eqnarray}
\tilde{G}_G({\bf r}_{12},E)=  \frac{1}{ab} \sum_{{\bf g}} 
\cos({\bf r}_{12}\cdot {\bf g})
\frac{\exp (Q[ 1-{\bf g}^2/(2E)])}{2E-{\bf g}^2} & r_{12}\neq 0
\end{eqnarray}
\[
-\frac{1}{4\pi}\sum_{{\bf \rho}} I(\frac{k}{2}|\rho+{\bf r}|)  \ \ ,
\]
The derivation of this expression require only slight
generalization of the derivation in ref.
\cite{BerrySin},
and we omit it. Note that the off diagonal terms are real.

To get a real expression 
for the determinant \EqRef{DeltaD}
vi simply extract a factor $d(E)$ from each row
\begin{equation}
\Delta_D(E)=(-d)^n det({\bf M})  \ \ ,
\label{eqn:DdM}
\end{equation}
where the matrix elements
\begin{equation}
M_{ij}= \left\{ \begin{array}{cc}
\frac{1}{4}\frac{Y_0(kR)}{J_0(kR)}+\tilde{G}_G^{(r)}(0,E) & i=j\\
\tilde{G}_G(r_{12},E) & i\neq j \end{array} \right.  
\label{eqn:Mij}
\end{equation}
are all real.

The energy dependence enter in the Green functions $\tilde{G}_G({\bf r},E)$
{\em and}
in the renormalized diffraction constant
\begin{equation}
\tilde{d}\equiv
\frac{1}{4}\frac{Y_0(kR)}{J_0(kR)}  \  \  ,
\end{equation}
where $k=\sqrt{2E}$. We will in computations artificially fix
$\tilde{d}$ and keep the energy dependence only in the Green functions, 
for reasons to be discussed later.

\subsection{Numerical considerations}
\label{s:num}

The numerical issue is to compute the Green functions with desired
accuracy. Through the Ewald summation technique each Green function is split
up into one sum over the dual lattice and one over the initial lattice $\sum I$.
There is however no closed expression for the function $I(x)$, 
introduced in
Eq. \EqRef{Ixdef}.
The asymptotic behavior of the function $I(x)$ is
\begin{equation}
I(x)=\frac{\exp(Q-x^2/Q)}{Q+x^2/Q}
    (1-\frac{x^2/Q-Q}{(x^2/Q+Q)^2}\ldots )  \ \ .
\end{equation}

As Berry noted, for sufficiently small $Q$ 
the sum $\sum I$ 
can be neglected, as far as the diagonal Green function
is concerned. This is not so for the off diagonal Green functions.
They are still small in absolute terms nevertheless significant.
We have chosen to include the sum $\sum I$, and compute the
integral \EqRef{Ixdef} by the asymptotic formula when appropriate. However, for
a small number of terms (a number decreasing with increasing energy)
the integral has to be evaluated numerically.
When $kr_{min}/Q\gg1$, where $r_{min}$ the smallest interdisk distance,
this is no longer an issue. This suggest a small value of $Q$.
One the other hand, a large $Q$ is preferred in the dual lattice sum, so the
choice of $Q$ is a compromise and can be adjusted according to energy.

The determinant was derived in the limit of small $kR$. The first 
correction will involve the factor $Y_1(kR)/J_1(kR)$ so our diffractive
determinant should work well whenever $kR\ll 1$.
For the $N=1$ case the first correction involved the factor $Y_4(kR)/J_4(kR)$,
the Bessel 
functions of order  1-3 are suppressed due to symmetry effects\cite{BerrySin}. 
The diffractive
approximation then works for slightly higher $kR \ll 4$. Indeed, we found in \cite{DahlVattay}
that $kR$ can be
rather close to the limiting $kR=4$ (for $N=1$) and presumably rather close to $kR=1$ in the
general case.

\section{Mean level density}
\label{s:mean}

In this section we will focus on the mean density of zeros of
the determinant $\Delta_D(E)$ as given in eqs \EqRef{DeltaD} and \EqRef{DdM}.
The density of states of 
the system in the diffractive limit is
asymptotically given by Weyls expression
\begin{equation}
\bar{d}_W=\frac{ab}{2\pi} \ \ .
\end{equation}
This density  does not need to be reproduced by 
the diffractive determinant $\Delta_D(E)$.
It will turn out that the
average density of zeros of $\Delta_D(E)$
will depend on the number of scatterers according to
\begin{equation}
\bar{d}^{zeros}_D=\frac{\min(N,4)}{4}\bar{d}_W  \ \ .  \label{eqn:bardD}
\end{equation}
That is, the 
full spectral density is achieved first when $N\geq 4$.
We will study spectra for fixed values of the parameter $\tilde{d}(kR)$ 
for reasons
to be discussed in the next section. 
But as the result \EqRef{bardD} does not depend
on the value of $\tilde{d}$, it will also apply to the (physical) case where
$\tilde{d}=\tilde{d}(kR)=\tilde{d}(\sqrt{2E}R)$ is allowed to vary with energy $E$.

Before actually deriving eq \EqRef{bardD} we will 
make some general comments.

The reason why we don't 
resolve the full density of states
for $N=1$ and $N=2$ has a simple
explanation in terms of symmetries
of the system.
The wave functions split up into the irreducible representations of the
respective group and our leading order determinant
cannot resolve them all.

If $N=1$ the symmetry is $C_{2v}$.
In a coordinate system with origin at the disk
 there is a reflection symmetry with respect to the
$x$ and $y$ axis. We only resolve 1/4 of the full spectral density,
namely those states with even parity with respect both axis.
To resolve the other subspaces one would need to take higher order terms
in the diffraction constant, 
and make a proper desymmetrization.
If $N=2$ there is a inversion symmetry with respect to the
point $({\bf r}_1+{\bf r}_2)/2$ and we recover 1/2 of the full spectral
symmetry.

It is obvious that for high enough $N$ 
one can avoid
this kind of symmetry effects.
It is also obvious that it is possible to construct 
configurations with large $N$
having a high degree of symmetry, like e.g. a regular lattice.
Below we will consider generic configurations from some random ensemble
for which there is no accidental symmetries of any kind.
We will also assume the absence of exact degeneracies in the
unperturbed spectrum.

Now to the derivation of eq \EqRef{bardD}.
First we show that the mean density of zeros of $\Delta_D$ equals the mean 
density of
poles of the same determinant. To show this can essentially repeat the 
arguments in ref \cite{BerrySin}. This was done for a slightly different determinant,
but the basic mechanism is the same.
The difference between the two integrated
spectral densities is given by \cite{BerrySin}.
\begin{equation}
\bar{N}^{zeros}_D(E)-\bar{N}^{poles}_D(E)=
-\frac{1}{\pi} \langle \mbox{Im} \; \log \Delta_D(E+i\epsilon) \rangle=
\end{equation}
\[
=-\frac{1}{\pi} \langle \mbox{Im} \; \log \;  \mbox{det} (1-{\bf T})\rangle
=-\frac{1}{\pi} \langle \mbox{Im} \; \mbox{tr} \; \log  (1-{\bf T})\rangle
\]
\[
=\frac{1}{\pi} \mbox{Im} \; \sum_{r=1}^{\infty}
\frac{\langle\mbox{tr} {\bf T}^r\rangle}
{r}  \  \  .
\]
A term $\mbox{tr} {\bf T}^r$ is just a product of Hankel functions and is zero on the mean.

Next we will compute the density of poles of $\Delta_D(E)$.
The poles will be located at the poles of the Green functions
whose density is $\bar{d}_W/4$. The problem is to determine
their multiplicity $m_N$.

So far, we have
\begin{equation}
\bar{d}^{zeros}_D=\bar{d}^{poles}_D=\frac{m_N}{4}\bar{d}_W 
\ \ .  \label{eqn:sofar}
\end{equation}

We now study the behavior of $\mbox{det}({\bf M})$ (with
${\bf M}$ defined in \EqRef{Mij}),
close to a pole corresponding to the quantum numbers $m,n>0$,
that is $2E$ is close to ${\bf g}^2=(2\pi)^2 ( (m/a)^2+(n/b)^2)$.
The matrix elements are then (approximately) given by
\begin{equation}
M_{ij}=\delta_{i,j}\tilde{d}+\frac{4}{ab}
\frac{\cos(2\pi m (x_j-x_i)) \cos(2\pi n (y_j-y_i))}
{2E-4\pi^2(m^2/a^2+n^2/b^2)}  \ \ ,
\end{equation}
where we have summed over the four (dual) lattice points 
$2\pi (\pm m/a , \pm n/b)$.
If we introduce the notation
\begin{equation}
\begin{array}{l}
\alpha_i=2\pi m x_i/a\\
\beta_i=2\pi n y_i/b
\end{array}  \ \ ,
\end{equation}
and
\begin{equation}
\lambda=-\frac{ab\tilde{d}}{4}2E-4\pi^2(m^2/a^2+n^2/b^2) \ \ ,
\end{equation}
the latter measuring the (small) distance to the pole,
we get
\begin{equation}
\mbox{det}{\bf M}=(\frac{\tilde{d}}{\lambda})^N 
\mbox{det} (
\lambda \delta_{i,j}-\cos(\alpha_j-\alpha_i)\cos(\beta_j-\beta_i))
\end{equation}
\[
\equiv (\frac{\tilde{d}}{\lambda})^N \mbox{det} {\bf \tilde{M}} \ \ .
\]
The rank of the matrix ${\bf \tilde{M}}$
is simply the requested multiplicity $m_N$
\begin{equation}
\mbox{det}{\bf M}=(\frac{\tilde{d}}{\lambda})^N 
O(\lambda^{N-\mbox{rank}({\bf \tilde{M}})})
\sim 1/\lambda^{\mbox{rank}({\bf \tilde{M}})}  \ \ .
\end{equation}
This rank is the maximum size a matrix having the structure
\begin{equation}
{\bf CC}_{ij}=\cos(\xi_j-\xi_i)\cdot \cos(\eta_j-\eta_i)  
\end{equation}
can have with a nonvanishing determinant.
To explore this problem we introduce
three other matrices
\begin{equation}
\begin{array}{r}

{\bf CS}_{ij}=\cos(\xi_j-\xi_i)\cdot \sin(\eta_j-\eta_i)\\
{\bf SC}_{ij}=\sin(\xi_j-\xi_i)\cdot \cos(\eta_j-\eta_i)\\
{\bf SS}_{ij}=\sin(\xi_j-\xi_i)\cdot \sin(\eta_j-\eta_i)\end{array}  \ \ .
\end{equation}
We further introduce the notation ${\bf CC}_j$ to mean the $j$'th column of
${\bf CC}$, and similarly for ${\bf CS}_j$ etc.
The idea is now to write column ${\bf CC}_j$  as the following linear
combination
\begin{equation}\begin{array}{ll}
{\bf CC}_j&=\cos(\xi_j-\xi_{j-1}) \cos(\eta_j-\eta_{j-1}){\bf CC}_{j-1}\\
       &   -\cos(\xi_j-\xi_{j-1}) \sin(\eta_j-\eta_{j-1}){\bf CS}_{j-1}\\
       &   -\sin(\xi_j-\xi_{j-1}) \cos(\eta_j-\eta_{j-1}){\bf SC}_{j-1}\\
       &   +\sin(\xi_j-\xi_{j-1}) \sin(\eta_j-\eta_{j-1}){\bf SS}_{j-1}
\end{array}  \ \ ,
\end{equation}
and similar relations hold for ${\bf CS}_j$, ${\bf SC}_j$ and ${\bf SS}_j$.
The result can be conveniently expressed in terms of matrices.
\begin{equation}
{\bf u}_j={\bf T}_j {\bf u}_{j-1}  \ \ ,
\end{equation}
where
\begin{equation}
{\bf u}_j= \left(
\begin{array}{l}
{\bf CC}_j\\{\bf CS}_j\\{\bf SC}_j\\{\bf SS}_j \end{array}
\right) \ \ ,
\end{equation}
and
\begin{equation}
{\bf T}_j={\bf T}(\xi=\xi_j - \xi_{j-1},\eta=\eta_j - \eta_{j-1}) \ \ ,
\end{equation}
and
\begin{equation}
{\bf T}(\xi,\eta)=\left(\begin{array}{cccc}
\cos(\xi)\cos(\eta) & -\cos(\xi)\sin(\eta)
    & -\sin(\xi)\cos(\eta) &\sin(\xi)\sin(\eta)\\
\cos(\xi)\sin(\eta) & \cos(\xi)\cos(\eta)
    & -\sin(\xi)\sin(\eta) &-\sin(\xi)\cos(\eta)\\
\sin(\xi)\cos(\eta) & -\sin(\xi)\sin(\eta)
    & \cos(\xi)\cos(\eta) &-\cos(\xi)\sin(\eta)\\
\sin(\xi)\sin(\eta) & \sin(\xi)\cos(\eta)
    & \cos(\xi)\sin(\eta) &\cos(\xi)\cos(\eta)
\end{array}\right) \ \ .
\end{equation}
Please note that the elements of ${\bf T}$ are scalars,
the elements of ${\bf u}$ are column vectors.
The external index $j$ has nothing to do with the internal structure
of these objects.
From the definition we have the relation
\begin{equation}
{\bf T}(\xi_j - \xi_{j-1},\eta_j - \eta_{j-1})
{\bf T}(\xi_{j-1} - \xi_{j-2},\eta_{j-1} - \eta_{j-2})
={\bf T}(\xi_j - \xi_{j-2},\eta_j - \eta_{j-2}) \ \ .
\end{equation}


The basic idea now is to explore whether it is
possible to write the column vector ${\bf CC}_n$
as a linear combination of the preceding columns  ${\bf CC}_j$ ($j<n$).
We address the corresponding problem for ${\bf CS}_n$, ${\bf SC}_n$ and ${\bf SS}_n$
simultaneously, and write
\begin{equation}
{\bf u}_n={\bf T}_n {\bf u}_{n-1}=
\mu_{n-1} {\bf u}_{n-1}+({\bf T}_n-\mu_{n-1} {\bf E}) {\bf u}_{n-1}
\ \ ,
\end{equation}
where $\mu_{n-1}$ is a multiplier and ${\bf E} $ is the unit matrix.
We carry on this procedure until we arrive at
\begin{equation}
{\bf u}_n=\sum_{j=2}^{n-1} \mu_j {\bf u}_j+{\bf S} {\bf u}_1 \ \ ,
\end{equation}
where
\begin{equation}
{\bf S}=({\bf T}_{n} {\bf T}_{n-1}\cdots {\bf T}_{2})
-\mu_{n-1} ({\bf T}_{n-1}{\bf T}_{n-2}\cdots {\bf T}_{2})-
\ldots \mu_2 ({\bf T}_{2}) \ \ ,
\end{equation}
or
\begin{equation}
{\bf S}={\bf T}(\xi_n-\xi_1,\eta_n-\eta_1)-
\sum_{j=2}^{n-1}\mu_j {\bf T}(\xi_j-\xi_1,\eta_j-\eta_1) \ \ .
\end{equation}
So the first $n$ columns of ${\bf CC}$ are linearly dependent
if and only if we can find multipliers such that
\begin{equation}\begin{array}{cc}
{\bf S}_{1j}=0 & j\neq 1 \end{array} \ \ ,
\end{equation}

The number of multipliers are $n-2$ and the number of equations to fulfill
is three.
So for generic parameters $\xi_i$ and $\eta_i$ the determinant of
${\bf CC}$ is zero for $n\geq 5$.
So $m_N=\mbox{rank}{\bf \tilde{M}}=\min(4,N)$.
Together with eq. \EqRef{sofar} the announced result \EqRef{bardD} follows.

\section{Level statistics}
\label{s:stat}

The disks are distributed randomly over the
torus according to a uniform distribution.
We compute spectra for individual members of this ensemble.
We choose the lattice constants to be $a=1$ and $b=2^{1/4}$,
the spectral statistics of the empty torus is than perfectly Poissonian;
exact degeneracies are avoided.

The critical parameter is $\tilde{d}(kR)$. As mentioned in the introduction,
statistical studies suffer for finiteness of the sample.
However if $R$ is sufficiently small a sufficiently large sample
can be obtained with essentially constant $kR$. But as we also demand that
$kR<1$ this would require too large values of $k$ to be numerically
tractable.
Instead we artificially fix $kR$ and and compute the bottom part of 
the spectrum.
We compute around 600 levels for each configuration - 
computations
do get a bit tedious for large $N$.

We are interested in two measures on the spectra, the 
integrated level spacing distribution $P(s)=\int_0^s p(s')ds'$
where $p(s)$ is the nearest neighbor spacing distribution.
Secondly we investigate the two point correlation function of levels
\begin{equation}
R(\epsilon)=
\langle \;
\sum_{ij} \delta((E-E_i)\bar{d}+\epsilon/2)\cdot
\delta((E-E_j)\bar{d}-\epsilon/2) \;
\rangle_E
\end{equation}
\[
=\delta(\epsilon)+\langle \;
\sum_{i\neq j} \delta((E-E_i)\bar{d}+\epsilon/2)\cdot
\delta((E-E_j)\bar{d}-\epsilon/2) \;
\rangle_E
\equiv
\delta(\epsilon)+\tilde{R}(\epsilon) \ \ ,
\]
where the average is taken for a large number of energies.
The correlation functions  are computed over a gaussian window
centered at the middle of the sample spectrum, 
its width is about  one sixth of the
sample size.
The results are then smeared with another gaussian having width $0.2$.

Below we will restrict our attention to spectral properties
of the cases $N=1$ and $N\geq 4$. The reader should bear in mind that for
$N=1$ only a quarter of the full density of states is resolved 
and the reported result apply to this single subspace.
Superposition of all four subspace would result in 
more Poisson-like statistics.

\begin{figure}
\epsffile{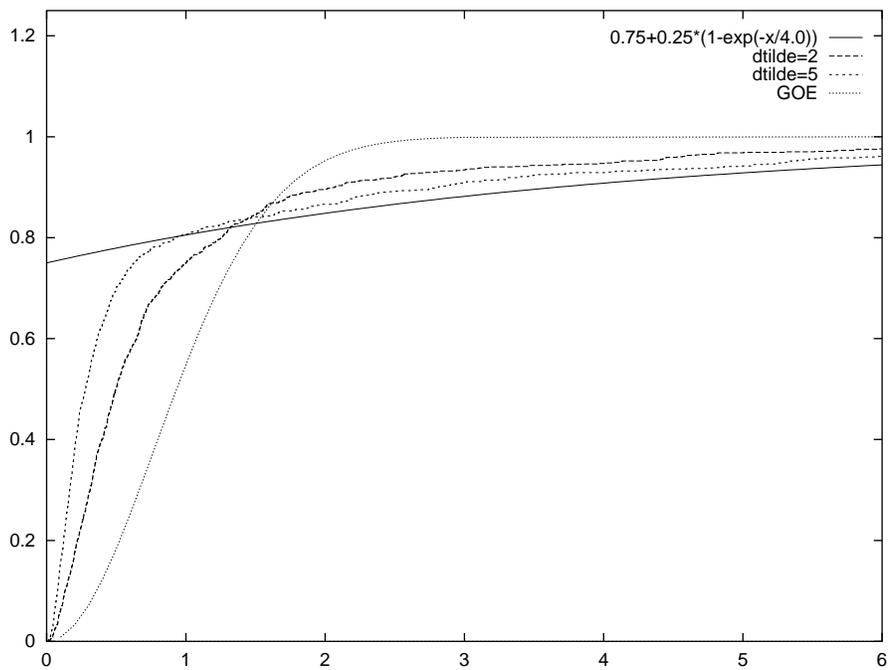}
\caption{The integrated level spacings distribution for
$N=7$ for two different values of $\tilde{d}$.}
\label{f:spac4push}
\end{figure}

The underlying spectrum of the empty torus reveals itself clearly in the
spectrum for large values of $\tilde{d}$.
If $N=1$ and $\tilde{d}=\pm \infty$ the spectrum is Poissonian for a trivial
reason, the zeros of the determinant have been pushed towards the
poles of the Green function corresponding to the spectrum of the
empty torus, cf. ref.~\cite{BerrySin}.
If $N\geq 4$ and
 $\tilde{d} \rightarrow \pm\infty$ four zeros will be pushed towards
each pole.
The corresponding limiting integrated level spacing distribution is then
\begin{equation}
P(s)=\frac{3}{4}+\frac{1}{4}\left( 1-\exp (-s/4)\right)  \  \  . \label{eqn:4push}
\end{equation}
This limiting distribution is plotted in Fig.~\ref{f:spac4push}
together with results
for two different values of $\tilde{d}$ ($N=7$).

\begin{figure}
\epsffile{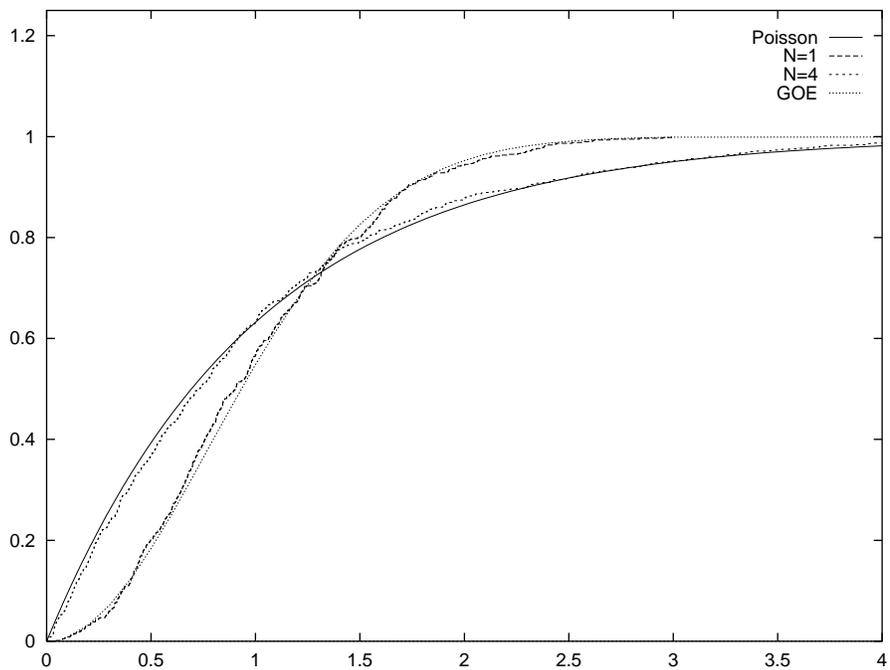}
\caption{The integrated level spacings distribution for
$\tilde{d}=0$ for  $N=1$ and $N=4$.}
\label{f:spac1and4}
\end{figure}

\begin{figure}
\epsffile{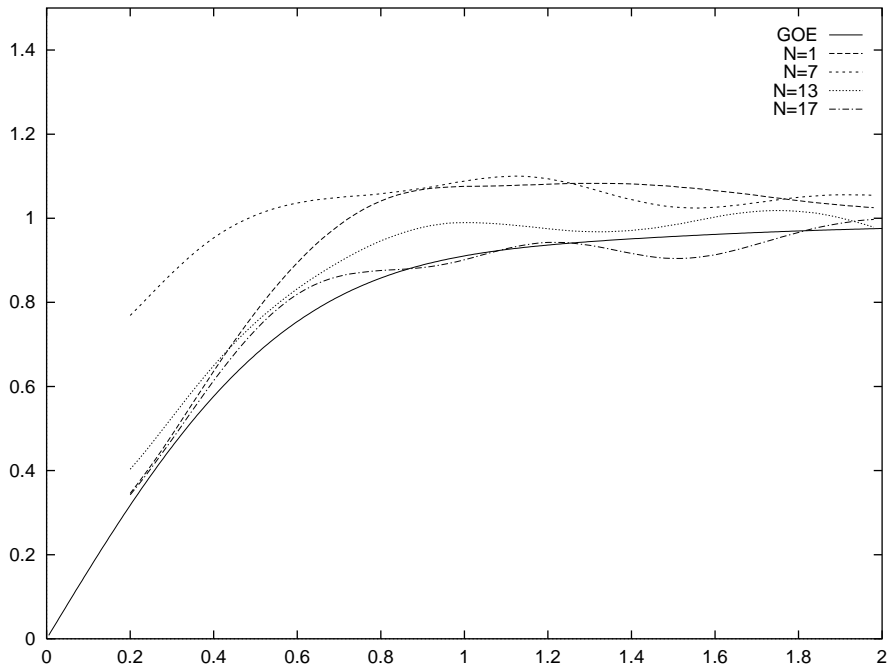}
\caption{The correlation function $\tilde{R}(\epsilon)$
for $\tilde{d}=0$ for variable
number of scatterers, compared with the GOE prediction.}
\label{f:tpcdpar0}
\end{figure}

For $N=1$ and $\tilde{d}=0$ the states are, so to say, repelled by the poles
of the Green function which result in a spectrum exhibiting level repulsion.
The level spacings distribution is very close to GOE, 
see \cite{DahlVattay} and Fig.~\ref{f:spac1and4}.
An exact agreement is not possible since the eigenvalues 
are locked between eigenvalues
of the integrable torus. This locking is released for high enough $N$
In fact, the two point correlation function $\tilde{R}(\epsilon)$ 
shows a clearer
deviation from GOE
than $P(s)$, see Fig.~\ref{f:tpcdpar0}.
One of the main questions is whether GOE can be approached as 
$N\rightarrow \infty$.

\begin{figure}
\epsffile{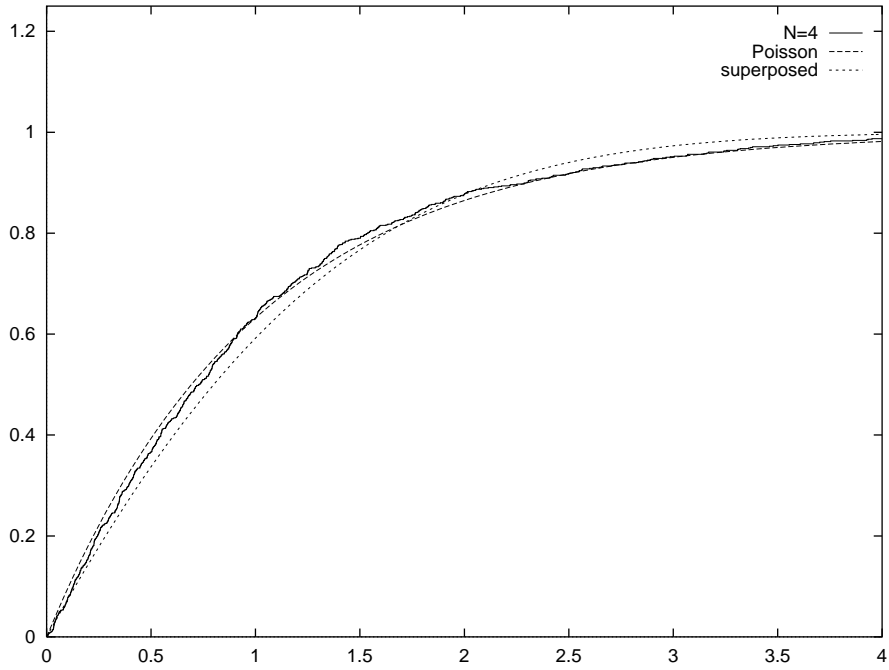}
\caption{The integrated level spacings distribution for
$\tilde{d}=0$, $N=4$ compared with the Poissonian spectrum and
the result from four superposed Wigner spectra.}
\label{f:4super}
\end{figure}

Increasing $N$ only to $N=4$, keeping $\tilde{d}=0$ (corresponding to
$kR=0.899\ldots$)
yields exactly the opposite result.
The level spacing distribution appears to be
perfectly Poissonian, see Fig.~\ref{f:spac1and4}.
It is known that Poisson-like distribution arises from independent
superposition of spectra, 
so one could think that that the determinant (for some unknown reason) 
factorizes.
However, the reported distribution
agrees better with the Poissonian prediction than 
with the statistics of
four superposed Wigner
spectra, see Fig.~\ref{f:4super}. 
So the statistics do appear to be Poissonian.
One could also reply that $kR=0.899\ldots$ is to close to $kR\approx 1$
to be physically relevant. However, we know that for $N=1$ the diffractive
approximation is very good close to $kR=0.899\ldots$ because a pole 
blows up the element in the KKR
matrix that corresponds to the diffractive approximation,
cf \cite{DahlVattay}, the same thing should happen if $N=4$, so we are 
probably talking
about a physical effect.

In Fig.~\ref{f:tpcdpar0} we 
keep $\tilde{d}=0$ constant and  increase
$N$ further. 
From now on we restrict our investigations to 
the correlation function $\tilde{R}(\epsilon)$.
According to the findings for $N=1$ we expect it to be a better indicator of deviations
from GOE.

We find that, indeed, the correlator
seems to approach that of GOE, for $N=13$ it already agrees better with GOE than
for $N=1$. 

\begin{figure}
\epsffile{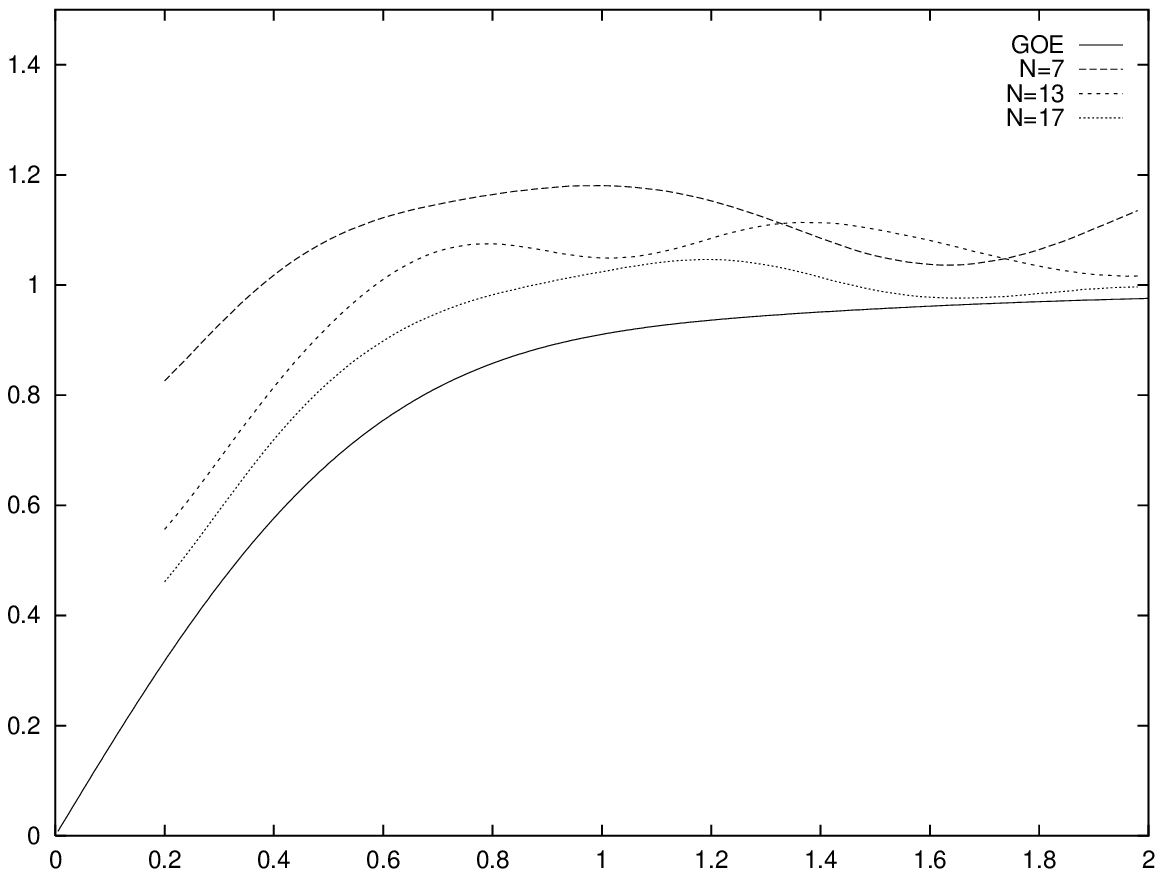}
\caption{The correlation function $\tilde{R}(\epsilon)$ for $\tilde{d}$ 
chosen according to eq \EqRef{dlog},
for variable
number of scatterers, compared with the GOE prediction.}
\label{f:tpclog}
\end{figure}

However, as we will see,
this result is not restricted too any particular choices
of $kR$.
Next we are going to consider another series of data. Suppose we are increasing
$N$ and at the same time decreasing $R$ in such a way that the  fraction of the
billiard area occupied by disks are kept constant:  $NR_N^2=C$. The
corresponding spectra are then studied in the neighborhood of some fixed $k$.
For small values of $kR$ we have\cite{AS}
\begin{equation}
\tilde{d}\equiv
\frac{1}{4}\frac{Y_0(kR)}{J_0(kR)} \approx \frac{1}{2\pi}
(\log (\frac{kR}{2})+\gamma) \ \ .
\end{equation}
We choose arbitrarily $k=2\exp{\gamma}/\sqrt{C}$ and thus
$\tilde{d}\approx -\log N/4\pi$,
and we are led to study the sequence
\begin{equation}
\tilde{d}_N=-\frac{\log N}{4\pi}  \  \  .  \label{eqn:dlog}
\end{equation}
The trend is the same, see Fig.\ \ref{f:tpclog}.
The correlation function approaches
that of GOE,but the approach is of course much slower.

The conclusions suggested by these studies are summarized in Table \ref{t:conc}.
The result in the lower right corner applies if the limit is approached
according to Eq. \EqRef{dlog} or something similar.
\begin{table}
\begin{tabular}{c|c|c|c|}
    & $|\tilde{d}|=0$ & $0<|\tilde{d}|<\infty$ & $|\tilde{d}|=\infty$\\
\hline
$N=1$ & $\approx$ GOE & & Poisson\\ 
\hline
$N=4$ & Poisson &   & 4$\times$ Poisson \\
\hline
$4<N<\infty$ & &  & 4$\times$ Poisson  \\
\hline
$N=\infty$ & GOE & GOE & GOE \\
\hline
\end{tabular}
\caption{}
\label{t:conc}
\end{table}

\section{Discussion}

The emergence of GOE in the limit of many small scatterers will hardly cause any big surprise.
The interesting thing is that the result has been studied within the framework of
periodic orbit theory. Admittedly, the periodic orbit was not used directly, that
would have led to unbearable slow convergence. We chose the 
underlying system in such a way that an efficient resummation could be
performed.

The similarity between the studied system and (disordered) antidot arrays
suggests that these can be successfully approached from the the diffractive limit
rather than from the Gutzwiller limit.
Higher order terms in the diffraction constant can be treated as perturbations.
A natural extension of the approach is to apply electric and magnetic fields
and study transport properties.

\vspace{1cm}
I am grateful to Gabor Vattay for
discussions and private lessons on the
geometric theory of diffraction. This work is a natural continuation of a joint project.
This work has been supported by the Swedish Natural Science
Research Council (NFR) under contract no.
F-AA/FU 06420-311.

\newcommand{\PR}[1]{{Phys.\ Rep.}\/ {\bf #1}}
\newcommand{\PRL}[1]{{Phys.\ Rev.\ Lett.}\/ {\bf #1}}
\newcommand{\PRA}[1]{{Phys.\ Rev.\ A}\/ {\bf #1}}
\newcommand{\PRD}[1]{{Phys.\ Rev.\ D}\/ {\bf #1}}
\newcommand{\PRE}[1]{{Phys.\ Rev.\ E}\/ {\bf #1}}
\newcommand{\JPA}[1]{{J.\ Phys.\ A}\/ {\bf #1}}
\newcommand{\JPB}[1]{{J.\ Phys.\ B}\/ {\bf #1}}
\newcommand{\JCP}[1]{{J.\ Chem.\ Phys.}\/ {\bf #1}}
\newcommand{\JPC}[1]{{J.\ Phys.\ Chem.}\/ {\bf #1}}
\newcommand{\JMP}[1]{{J.\ Math.\ Phys.}\/ {\bf #1}}
\newcommand{\JSP}[1]{{J.\ Stat.\ Phys.}\/ {\bf #1}}
\newcommand{\AP}[1]{{Ann.\ Phys.}\/ {\bf #1}}
\newcommand{\PLB}[1]{{Phys.\ Lett.\ B}\/ {\bf #1}}
\newcommand{\PLA}[1]{{Phys.\ Lett.\ A}\/ {\bf #1}}
\newcommand{\PD}[1]{{Physica D}\/ {\bf #1}}
\newcommand{\NPB}[1]{{Nucl.\ Phys.\ B}\/ {\bf #1}}
\newcommand{\INCB}[1]{{Il Nuov.\ Cim.\ B}\/ {\bf #1}}
\newcommand{\JETP}[1]{{Sov.\ Phys.\ JETP}\/ {\bf #1}}
\newcommand{\JETPL}[1]{{JETP Lett.\ }\/ {\bf #1}}
\newcommand{\RMS}[1]{{Russ.\ Math.\ Surv.}\/ {\bf #1}}
\newcommand{\USSR}[1]{{Math.\ USSR.\ Sb.}\/ {\bf #1}}
\newcommand{\PST}[1]{{Phys.\ Scripta T}\/ {\bf #1}}
\newcommand{\CM}[1]{{Cont.\ Math.}\/ {\bf #1}}
\newcommand{\JMPA}[1]{{J.\ Math.\ Pure Appl.}\/ {\bf #1}}
\newcommand{\CMP}[1]{{Comm.\ Math.\ Phys.}\/ {\bf #1}}
\newcommand{\PRS}[1]{{Proc.\ R.\ Soc. Lond.\ A}\/ {\bf #1}}


\newpage

\end{document}